\documentclass{article}

\usepackage{arxiv}

\usepackage[utf8]{inputenc} % allow utf-8 input
\usepackage[T1]{fontenc}    % use 8-bit T1 fonts
\usepackage{hyperref}       % hyperlinks
\usepackage{url}            % simple URL typesetting
\usepackage{booktabs}       % professional-quality tables
\usepackage{amsfonts}       % blackboard math symbols
\usepackage{nicefrac}       % compact symbols for 1/2, etc.
\usepackage{microtype}      % microtypography
\usepackage{lipsum}		% Can be removed after putting your text content
\usepackage{graphicx}
\usepackage{natbib}
\usepackage{doi}
\usepackage{amsmath}

\usepackage{xcolor}
\usepackage{listings}
\usepackage{tikz}

\title{Empirical Network Structure of Malicious Programs}

\author{ \href{https://orcid.org/0000-0001-7646-4328}{\includegraphics[scale=0.06]{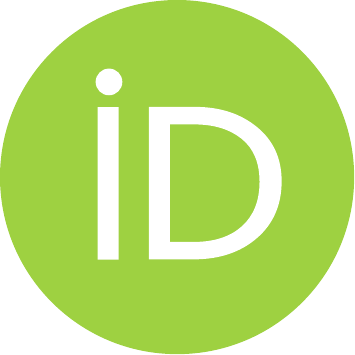}\hspace{1mm}John Musgrave}\thanks{Computer Science Department, Northern Kentucky University, Highland Heights, KY} \\
	Department of Computer Science\\
	University of Cincinnati\\
	Cincinnati, OH \\
	\texttt{musgrajw@mail.uc.edu} \\
	\And
	Alina Campan \\
	Department of Computer Science\\
	College of Informatics\\
	Northern Kentucky University\\
	Highland Heights, KY \\
	\texttt{campana1@nku.edu} \\
	\And
	Temesguen Messay-Kebede \\
	Air Force Research Lab\\
	Wright-Patt Air Force Base\\
	Dayton, OH \\
	\texttt{temesgen.kebede.1@us.af.mil} \\
	\And
	David Kapp \\
	Air Force Research Lab\\
	Wright-Patt Air Force Base\\
	Dayton, OH \\
	\texttt{david.kapp@us.af.mil} \\
	\And
	Anca Ralescu \\
	Department of Computer Science\\
	University of Cincinnati\\
	Cincinnati, OH \\
	\texttt{ralescal@ucmail.uc.edu} \\
}

\date{August 20, 2022}

\renewcommand{\headeright}{}
\renewcommand{\undertitle}{}
\renewcommand{\shorttitle}{\textit{arXiv} Template}

\hypersetup{
pdftitle={Empirical Network Structure of Malicious Programs},
pdfsubject={malware analysis, graphs, network science, security},
pdfauthor={John Musgrave, at al.},
pdfkeywords={malware analysis, graphs, network science, security},
}

\begin{document}
\maketitle

\begin{abstract}
A modern binary executable is a composition of various networks.  Control flow graphs are commonly used to represent an executable program in labeled datasets used for classification tasks.  Control flow and term representations are widely adopted, but provide only a partial view of program semantics.  This study is an empirical analysis of the networks composing malicious binaries in order to provide a complete representation of the structural properties of a program.  This is accomplished by the measurement of structural properties of program networks in a malicious binary executable dataset.  We demonstrate the presence of Scale-Free properties of network structure for program data dependency and control flow graphs, and show that data dependency graphs also have Small-World structural properties.  We show that program data dependency graphs have a degree correlation that is structurally disassortative, and that control flow graphs have a neutral degree assortativity, indicating the use of random graphs to model the structural properties of program control flow graphs would show increased accuracy.  By providing an increase in feature resolution within labeled datasets of executable programs we provide a quantitative basis to interpret the results of classifiers trained on CFG graph features.  An increase in feature resolution allows for the structural properties of program classes to be analyzed for patterns as well as their component parts.  By capturing a complete picture of program graphs we can enable theoretical solutions for the mapping a program's operational semantics to its structure.
\end{abstract}

% keywords can be removed

\section{Introduction}
In this study we propose a quantitative analysis of program networks.  The use of structured feature representations in program networks can increase feature resolution and are directly correlated to a program's operational semantics.

\subsection{Related Work}
Machine learning techniques have been applied in many contexts to successfully identify malicious programs based on a variety of features.  Many classification methods have been used for supervised learning including deep neural networks and support vector machines.  Several datasets have been collected with various kinds of features, including assembly instructions, n-gram sequences of instructions and system calls, and program metadata  \cite{souri2018state}, \cite{rawashdeh2021single}, \cite{kebede2017classification}, \cite{djaneye2019static}, \cite{chandrasekaran2020context}.

A number of studies have explored the use of static features at the level of file format, and their impact on the classification of malicious programs.  Decision trees for the classification of Windows PE files have shown to be effective in classifying malicious programs.  Subsequent studies have focused on malware classification using ensemble methods, which include random forest with support vector machines and principal component analysis that focused on features extracted from file headers in Trojan malware \cite{shafiq2009pe}, \cite{siddiqui2008detecting}, \cite{witten1999weka}.

The focus of many studies applying machine learning techniques to malware analysis is the task of classification for the purposes of identifying unknown programming errors.  Zhou et al. used a graph neural network (GNN) model to classify various types of C functions in order to determine semantic errors in their abstract syntax tree (AST) and program flow.  The model was trained on a dataset of functions which were drawn from several executable binaries including the Linux kernel.  Wang et al. developed a synthetic dataset of 3 million Python programs with class labels, and extracted function call graphs from AST graphs generated from tokenization to be used for training with a novel GNN design.  Park et al. have used sequence modeling to identify potential optimizations in programs based on an intermediate representation.  A program flow graph was extracted from this intermediate program representation and used in the sequence predictions \cite{zhou2019devign}, \cite{wang2020learning}, \cite{park2012using}.

Several studies have used control flow graphs as features in datasets used for malware classification tasks.  Bruschi et al. have extracted control flow graphs from malware for the purposes of classification through comparing the graphs for isomorphism.  Cesare et al. have presented several studies on the uses of control flow graphs in the classification of malware with efficient results \cite{bruschi2006detecting}, \cite{cesare2010fast}, \cite{cesare2013control}, \cite{cesare2010classification}.

\subsection{Current Work and Motivations}
Machine learning methods of malware analysis are widely claimed to represent features of operational semantics through models and classification accuracy.  The results and accuracy of classification techniques are dependent on the feature representations used in these datasets.  Features are extracted from datasets collected at specific levels in an architectural hierarchy.  Many useful features for classification can be extracted at multiple points in the architectural hierarchy as discussed in previous studies, e.g. instruction n-grams, sequences and patterns of bytecode or hex representations, as well as graphs, n-grams, and sequences of system API calls.  The feature representation selected determines the granularity available and the degree to which classification accuracy is correlated with and representative of semantics.  However, analysis of malicious programs presents several obstacles for an accurate classification of programs based on their operational semantics.  Class labels are often coarse grained, with one label representing the class of an entire program, without a clear method to provide increased resolution for supervised models which are dependent upon labeled data.  Without an increased resolution of features that are descriptive of structure, explaining correlation between structure and semantic abstraction is very challenging.  The degree to which a program's component parts contribute to a class cannot be determined without increased feature resolution in a labeled dataset.  A program level of resolution is too low to provide meaningful information about the relationship between a class label and a program's operational semantics across abstraction layers.  Therefore we view feature representations from two perspectives:  as a description of program operational semantics, and the ability to describe structural properties.  Structural properties enable the syntactic elements of a program to be interpreted.  Semantic properties allow the correctness of a program to be verified across abstraction layers.  The relationship between syntax and semantics in natural language has been successfully modeled by using topics in bi-partite networks  \cite{sebesta1999programming}, \cite{souri2018state}, \cite{griffiths2007topics}.

Accurate methods of analysis and proof of program semantics require the use of a program specification to be checked for validity.  Malicious programs do not have a specification available for verification prior to execution.  Without the presence of a formal specification, proofs and verification for the operation of a program cannot easily be developed.  Therefore, a semantic representation corresponding to a formal specification or other description of operational semantics must be constructed from structural elements to verify the semantic correctness.  This method of construction represents a bottom up approach.  The degree to which an abstract representation correlates to a program's semantics is an open question that has yet to be answered. 

In order to explain how structural elements are correlated to their abstract semantics, the structural properties must first be defined.  Without the empirical observation of structural properties, the patterns in structure cannot be identified, and therefore the semantic abstraction generating structural patterns cannot be derived.  A definition of semantics must be specified, and in this context we refer to the operation of a program.  Program semantics that are descriptive of the behavior of a program must persist across architectural layers.  While other semantic representations exist at various architectural levels, a structured feature representation and analysis of instructions composed of opcodes and their operands does not require proving a correlation between architectural levels.  The instructions describe the operation of the machine at the most fundamental level.  In the case of malware, a malicious binary is the initial artifact.  While we are unable to make assumptions about the values operands will have at runtime, we are able to derive the structure of their dependency.  So when we discuss structure in a program, we are describing a structure of dependencies.  This is explicitly specified in the program, and we cannot determine the values of the terms at runtime.  By describing the structure of dependency we obtain a representation that is descriptive of a program's structural properties.  By representing program structure at the lowest level of abstraction we directly describe the structural properties of program operation in terms of dependency.  So the ability to recognize structural patterns that are tied to semantics would have wide applications.

Many studies using machine learning methods to classify malware focus on finding errors in high level languages.  While this is useful for increased security in the automation of software development, it does not address the semantic interpretability of the classification results.  Classification does not directly provide a fine grained description of the malicious program structure, and is dependent on the granularity of feature representation used for training.  Features in labeled datasets are often at the most coarse grained level of the binary as a whole.  Studies based on malware graph features often have a focus on methods of differentiation in control flow graphs through graph isomorphism.  While this is a useful feature for the classification of malicious programs, it is only a partial view of the program's operational semantics.  The isomorphism of graphs and all subgraphs is used to determine class equivalence, and have not been directly measured for structural properties.  Further, if the classification model was trained on a high level language such as C or Java, a correlation is required to be proven to the compiled artifact.  The correlation of the structural properties identified through classification to an abstract semantic representation is an open question.  We attempt to analyze structural properties of program networks that correlate directly to operational semantics \cite{bruschi2006detecting}, \cite{arora2012evaluation}.

Why should we measure program control flow?  When an executable program is viewed in automata theoretic terms as the operation of a Turing Machine, then the state of the Finite State Machine is subject to analysis.  The potential state space of a program is not computationally feasible to analyze, and presents challenges for searching within this space.  However, the set of state transitions and structure of their dependencies within the same program contain the program's structure and are subject to further analysis to gain insight.  This is the structure that provides insight into the sequences present in the program. Network analysis methods provide meaningful features for this analysis.  These features are needed for accurate classification tasks based on program behavior \cite{hopcroft2001introduction}, \cite{cesare2013control}. 

The instruction sequence in the program is also not determined by the linear placement of the term in the document, but by the structure of the program's control flow network.  This sequence is also segmented into blocks of potentially deterministic sequences of instructions \cite{hopcroft2001introduction}. 

Why is data movement important?  Since programs are made up of sequences of instructions, by performing an analysis of the frequency of terms in the sequence of instructions using $tf-idf$ based methods, the result shows an overwhelming prevalence for a high frequency of data movement instructions.  This drastic distribution of term frequencies shows the highest density in the body of the distribution, with the tail being extremely thin.  In order to further analyze the body of the term distribution, a feature set with more structure is required for meaningful analysis.  The underlying semantic structure is not captured by the term frequency distribution of the instructions alone.  Networks of data dependencies however are descriptive of the structural relationships between the terms.  This network structure is directly descriptive of the terms present in the body of the term frequency distribution represented by $tf-idf$, which is positively skewed.  Our study measures network properties at a segment and program level for data movement and program execution respectively \cite{musgrave2020semantic}. 

The goal of this work is to perform quantitative analysis of the networks which compose malicious executable programs and to empirically describe the properties of the networks’ structure.  This is for the purposes of providing a method of collecting a set of structured features for future use in the representation and classification of malware operational semantics.  In the absence of a formal specification, semantics are represented by patterns in structural properties of the program.  Networks provide such a representation, and can be analyzed for their structure.  Constructing networks from the lowest architectural level is a direct representation of the instructions being executed, and does not require translation across architectural layers.  The overarching goal is further depth of abstraction in the semantic representations of malicious programs for increased accuracy in malware classification tasks.

Several approaches exist to analyze a program based on its behavior, including static and dynamic analysis, or collecting execution traces $a-posteriori$.  In the case of malware analysis, a formal specification for a program does not exist.  A binary executable is the sole artifact for analysis. For this reason it is necessary to take a bottom up approach to the structural analysis, rather than collect artifacts of higher level language descriptions.  Further, the class of non-minimal equivalent automata for a given deterministic finite automata is infinite.  Therefore a binary executable may have a very large amount of high level language representations that are semantically equivalent \cite{hopcroft2001introduction}. 

At this time we are unaware of a study that directly measures the structural properties of the program networks.  Many studies have successfully trained classifiers using Graph Neural Networks and Graph Convolutional Neural Networks.  In some cases these studies have extracted program control flow graphs and compiled labeled datasets for models trained on graph features, but the identification of the structural properties of the networks was not the focus, so the measurements are not known.

The power law distribution of network degrees determines what statistical tools are applicable.  In order to correlate networks by their degree, we compare networks based on their assortativity.  As we will see, networks' structural assortativity properties for the collected data appear to hold across samples.  This means that predictions can be made as to the degree correlations for various networks.  The structure of data dependency graphs can be predicted to not have a prevalence of links between nodes with high degree.   The structure of control flow graphs can be predicted based on the network properties of random graphs.  The values of the random graph properties will likely vary by sample, and this is an area of future research.  

\subsection{Outline}
Section 2 covers the experiments performed.  Section 3 contains the Results and Discussion.  Section 4 is a Summary and Conclusion.

\section{Experiments}
This section describes the data collection process as well as the metrics selected for our analysis.

\subsection{Data Collection}
The goal of this study is to provide a quantitative basis for analysis of the structure of malicious executable programs.  The executable programs for our purposes are adversarial, and are provided in binary form.

Several feature representations exist in the binary, and additional data can be collected for more structured representations.  For example, a program can be viewed as a document in a $tf-idf$ representation, with terms selected from a dictionary.  Terms in a dictionary correspond to assembly instruction opcodes, which are explicitly specified the binary.  However, this ignores any data operands, and focuses on the term frequency distribution as the primary representation. Also, unless further structure is considered, this assumes a linear structure to the document.  Executable programs are not structured linearly, but are divided into segments, which are structured in a network.  This is one motivation for a structural analysis of the networks present within a program.

The $tf-idf$ representation makes two assumptions prior to analysis, that data movement operations should be ignored, and that the distribution of term frequencies is representative of a program.  However, the term distribution is heavily skewed towards the use of data movement instructions, and has a tail that is thin.  The variance captured by data movement in the body is more than the variance of all other terms in the tail. So ignoring data movement operations discards a majority of the data, and this data is descriptive of the program's function.  Further, we are unable to further analyze the body of the term distribution without additional quantitative information about the structure \cite{souri2018state}.

Since our use case focuses on adversarial examples, we have only selected malicious examples to be analyzed for their program structure.  No specifications exist beforehand for verification for these samples.  Additionally, each malware example must be able to be executed as a binary.  The program samples were selected from the public malware repository $theZoo$, a collection of live malware samples. We plan to expand to other publicly available malware repositories in future studies \cite{thezoo}.

Each sample was taken from a binary able to be executed on the target platform across several operating systems and architectures.  Each binary was decompiled using GNU $objdump$, a tool which is able to reverse engineer a binary program to its assembly instruction set representation. Assembly instruction representations were collected for each program in the dataset.  Our assembly artifacts were segmented into basic blocks, sequential segments of contiguous instructions separated by a jump instruction.  Control flow graphs are obtained from static and dynamic analysis tools, and are represented in an adjacency matrix format.  We used several tools for comparison purposes, but focus primarily on $radare2$, although some variation exists between the tools, and we do not offer a comparison of program static analysis tools in terms of their accuracy.  The graphs recovered by static analysis tools are obtained by analyzing the structure of basic blocks as nodes in a program networks, and jump instructions as edges to these nodes.  One control flow graph exists for each program in our dataset \cite{nar2019analysis}.

Since basic blocks determine the nodes in the program network, each node was analyzed for its data dependency structure.  Each line in the segment is an assembly instruction composed of an opcode and a set of operands.  Since these instructions are issued with a corresponding order, explicit and implicit data dependencies exist between the instructions in the sequence.  Data dependency graphs were constructed by creating a network where nodes represent a data operand, and an edge between the nodes represents a $mov$ instruction, or other opcode with implicit data movement between source and destination operands.  This was done by using a program written to look for specific data operands of instructions in the same basic block being input, and output an adjacency list representing the data dependency graph.  The graph of dependencies between operands in assembly instructions was constructed for each basic block in the program, and were represented in an adjacency list format for each graph.  Figure 1 shows an example of a data dependency graph being constructed from a sequence of contiguous assembly instructions with dependencies in a basic block.  We constructed data dependency graphs for each block, and focused solely on dependencies between $mov$ instructions, as the prevalence of data movement was the primary motivation for providing additional structure \cite{hennessy2011computer}, \cite{hagberg2008exploring}.

The interaction of the two graphs provides a large amount of additional data that can be used for further analysis.  This interaction is captured by a program dependence graph.  The program dependence graph can be obtained by constructing a bi-partite graph from a tensor representation, where each cell in the tensor represents a data dependency graph, and the overall tensor structure is built from the adjacency matrix for the program's control flow graph.  We have constructed a program dependence graph in a tensor representation, which we include as a note.  A program dependence graph represents a composition of the networks analyzed, and does not differ in the structural properties of its components.  A complete semantic analysis of the PDG composition is outside the scope of this work, but we present a structural analysis of its component parts \cite{ferrante1987program}.

Therefore three networks are available for analysis for each program sample, the control flow graph (CFG), the data dependency graphs (DDG) for each node, and the complete program dependence graph (PDG).  The PDG represents the interaction of the CFG and DDG graphs.  Each of the graphs collected are directed graphs which contain cycles, and can also be analyzed as undirected graphs \cite{hennessy2011computer}, \cite{ferrante1987program}.

The quantitative network properties discussed in the results section were observed by using network libraries to measure the adjacency matrix and adjacency list representations of the networks collected.  Additional analysis was performed in Matlab to obtain quantitative properties of the adjacency matrices and generate plots of the data \cite{hagberg2008exploring}, \cite{toolbox1993matlab}.

\subsection{Network Metrics}
We briefly introduce for background several metrics for our analysis that are commonly used in the relevant literature \cite{barabasi2013network}.

$N$ represents the number of nodes in the network, or the size of the network.

$L$ represents the number of edges or links in the network, directed edges in the context of a directed graph.

$K$ represents the degree of a given node in the network, calculated by counting the number of links for a given node.

$k_{max}$ represents the maximum degree for a node in the network.

$k_{max} / ln(N)$ - the ratio of a network's maximum degree and the natural log of the network size in terms of nodes.  This measurement is a predictor of network diameter, and also indicates cluster size in Small-World networks.  

$\gamma$ is a term that represents the power law exponent that the degree distribution follows, when present.

\subsection{Scale-Free}
Scale-Free networks have a number of interesting properties including the generative property of preferential attachment.  For the purpose of this study, the most relevant feature is the presence of a degree distribution following a power law exponent that is sufficiently large to cause a hub and spoke pattern \cite{barabasi2013network}.

\subsection{Small-World}
A Small-World property of a network is characterized by a small network diameter and a high average clustering coefficient.  In order to demonstrate the Small-World property, we use $k_{max} / ln(N)$ as a measurement of network diameter over $ln \ N / ln < K >$, although we present both measurements for comparison.  Since our degree distributions follow a power law distribution we use the former metric to measure the existence of the Small-World property as it is not dependent on mean degree.  Both represent predictions of network diameter \cite{watts1998collective}. 

\subsection{Degree Assortativity}
Degree assortativity is the correlation of nodes with high degree.  This is representative of the existence of a trend for nodes with high degree to have links between them, or not.  Networks with nodes that have high degree that show a preference to link together over low degree nodes are assortative.  Networks with high degree nodes that show a preference not to link together, and prefer to link to low degree nodes are disassortative.  Networks that do not show a preference among high degree nodes are neutrally assortative.

\begin{figure}[t]
    \centering
    \hspace{5cm}
    \begin{lstlisting}
                            mov    ecx, rbp - 44
                            mov    eax, ecx
                            and    eax, 400
                            or     eax, 140
                            or     ecx, 1
                            cmp    rip + 170, 0
                            cmovne ecx, eax
                            mov    rbp - 44, ecx
                            mov    rip + 180, 0
                            jmp    0x100000000
    \end{lstlisting}
    
    \vspace{1cm}
    
      \begin{tikzpicture}[node distance={15mm}, thick, main/.style = {draw, circle}] 
    \node[main] (1) {$a_1$}; 
    \node[main] (2) [right of=1] {$a_2$}; 
    \node[main] (3) [below of=1] {$a_3$}; 
    \node[main] (4) [right of=3] {$a_4$}; 
    \draw[->] (2) -- (1); 
    \draw[->] (1) -- (3); 
    \draw[->] (3) -- (1); 
    \draw[->] (1) -- (4);
    \draw[->] (4) -- (1); 
    \end{tikzpicture}
    \[ A_{ddg} = \{ a_i \ | \ a_i \ \in \ A_{operand} \} \]
    \caption{Basic block segment of assembly instructions and its data dependency graph.  The data dependency graph shown is constructed from data movement instruction dependencies.  $mov$ instructions are the primary instructions with respect to term frequency.}
\end{figure}

\begin{figure*}[t]
    \hspace{-3.5cm}
    \includegraphics[scale=0.4]{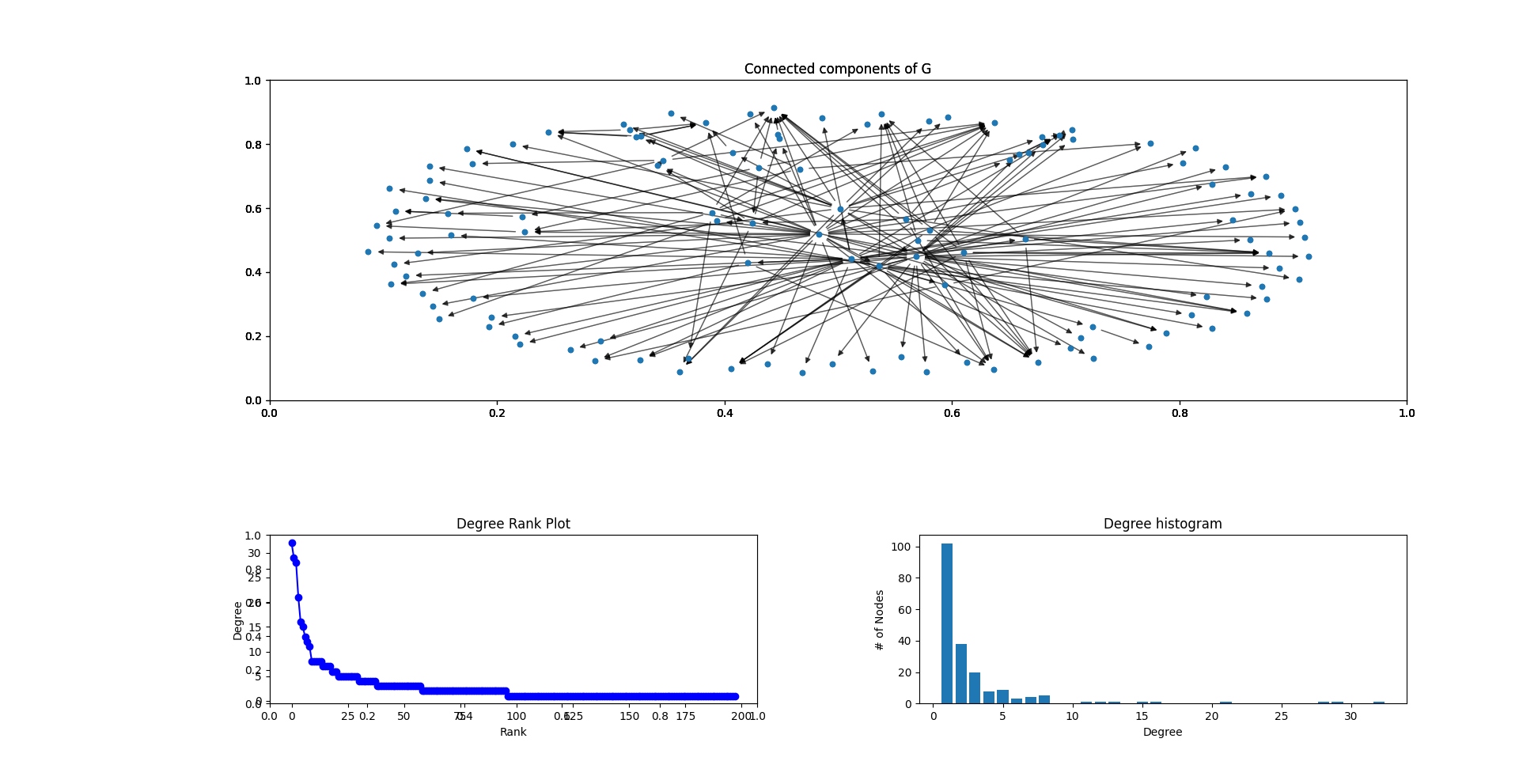}
    \caption{\textit{The program's Control Flow Graph has a power law degree distribution} -  Figure showing a network with Degree Histogram, and Degree Rank Plot of a program's Control Flow Graph (CFG), which shows a power law degree distribution with a positive skew, in that most nodes have very few connections and a small number of nodes have a high degree.}
\end{figure*}

\section{Results}

\begin{figure*}[t]
    \centering
    \hspace{-2cm}
    \begin{tabular}{||c| c c c c c c c||} 
\hline
Malware Block Index & $N$ & $L$ & $k_{max}$ & $k_{max}/ln(N)$ & $ln(N)/ln<K>$ & $Pearson$ & $\gamma$\\ [0.5ex] 
\hline\hline
\hline
390 & 48 & 32 & 3 & 0.774 & 2.904 & -0.153 & 18.264 \\
\hline
527 & 40 & 30 & 9 & 2.466 & 2.397 & -0.416 & 4.538 \\
\hline
263 & 29 & 26 & 4 & 1.187 & 1.878 & 0.105 & 3.281 \\
\hline
358 & 32 & 25 & 5 & 1.442 & 2.218 & -0.577 & 4.279 \\
\hline
526 & 21 & 13 & 4 & 1.313 & 2.459 & -0.326 & 8.574 \\
\hline
    \end{tabular}
    \caption{Comparison of network structure between data dependency networks of operands for largest 5 DDG networks.}
    
    \vspace{0.05cm}
    \hspace{-4cm}
    \includegraphics[scale=0.65]{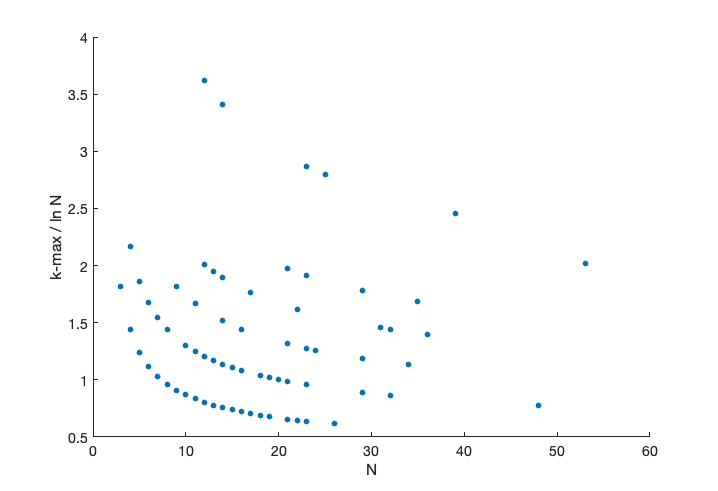}
    \caption{Scatter plot of network size and ratio of natural log of maximum node degree and natural log of network size for data dependency networks collected for a single program sample.  Network diameters are small, and decrease as network size increases, additionally indicating the existence of network hubs. }
\end{figure*}

\begin{figure}[t]
    \hspace{-2cm}
    \includegraphics[scale=0.34]{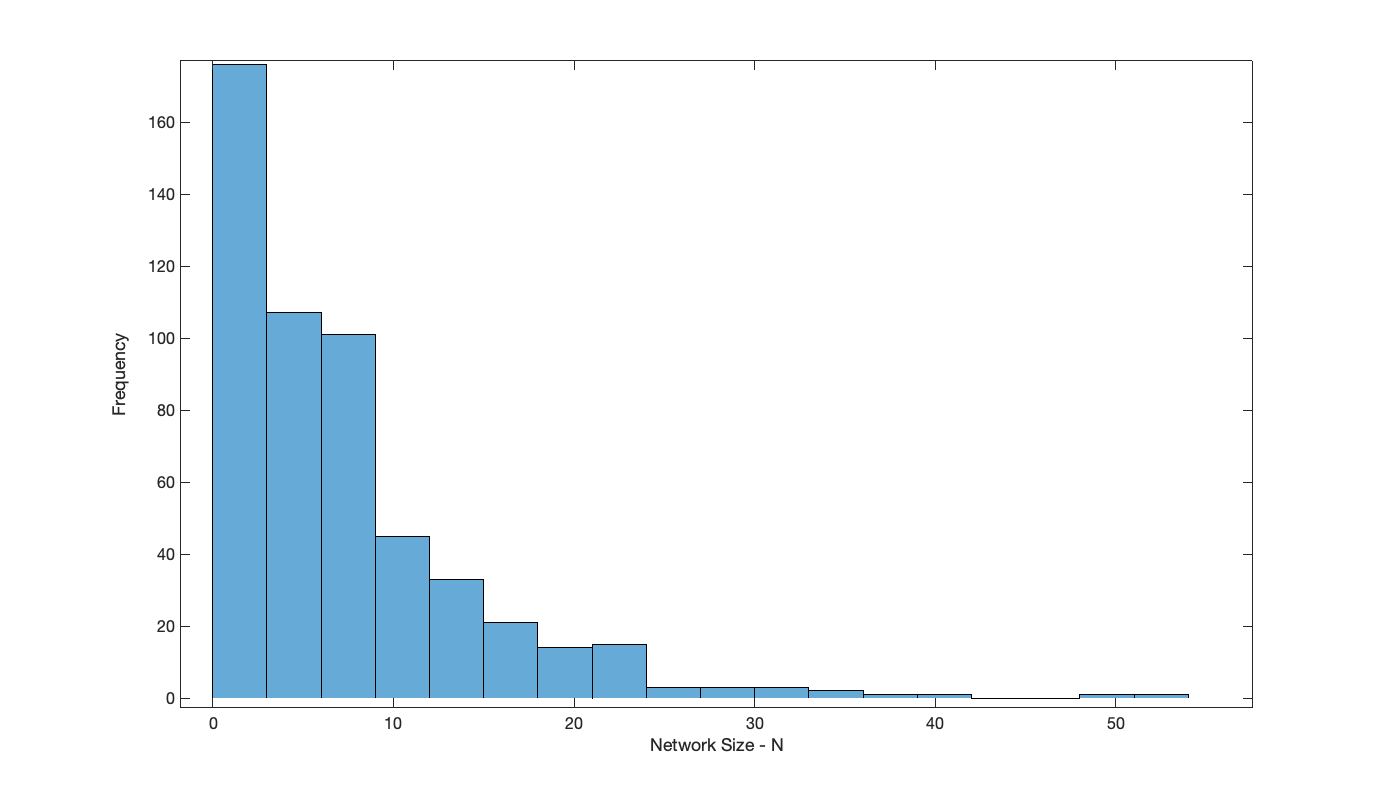}
    \caption{\textit{Data Dependency Graph sizes for this program are skewed positively and follow a power law distribution} - This histogram shows network sizes $N$ of data dependency networks (DDG) extracted from $mov$ instructions per block segment in a single program.  This shows a power law distribution where most data dependency networks for $mov$ instructions are very small, $N$ less than 5.}
\end{figure}

\begin{figure*}[t]
    \centering
    \hspace{-4cm}
    \includegraphics[scale=0.63]{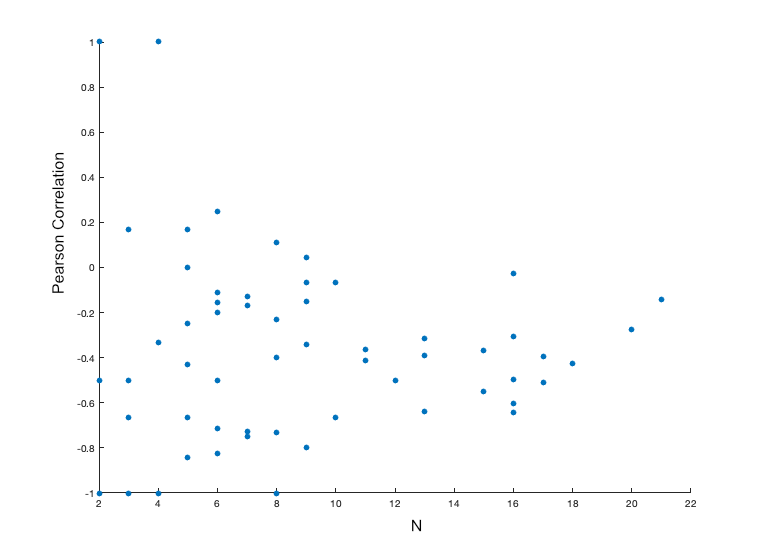}
    \caption{\textit{Data Dependency Networks' Degree Correlations are Disassortative and are not random for the sample} - Scatter plot of network size $N$ on the x-axis and Pearson correlation of the network in terms of degree on the y-axis for data dependency networks of $mov$ instructions, $DDG$, per block segment.  This shows that the degree correlation coefficient for a majority of networks is below 0, meaning that a majority of the data dependency networks in this program are Degree Disassortative, and do not link to nodes with high degree.}
\end{figure*}

As discussed in the experiments, Figure 1 outlines the construction of data dependency graphs from basic block segments.  These graphs were constructed for dependencies of data operands between data movement instructions, which were the primary motivation for the analysis of graphs.  Data movement instructions make up the most significant portion of the term variance with respect to frequency.

Figure 2 shows various measurements taken from program's control flow graph.  The network's degree histogram is shown in the bottom right, which is positively skewed with the largest number nodes in the network having a low degree.  The topology of connected components is shown in the top of the figure, which shows several nodes with few connections, and a small number of nodes with high degree.  The network's degree rank plot is shown in the bottom left.  The control flow graph for a program represents a network of contiguous program instructions as nodes with transitions between them as edges.  The degree histogram shows that the degree distribution follows a power law, a small number of nodes have a very high node degree.  Figures 4 and 7 both show the comparison of $k_{max} / ln(N)$ with $ln(N)/ln < K >$ for comparison purposes.  While degree distributions of control flow graphs follow a power law, this distribution is not stable within a program.  The degree-rank plot in Figure 2 shows a power law trend, but when this power law distribution is plotted on a log scale, the cumulative distribution function shows that the power law exponent decays from the linear trend.  This shows that the power law exponent of the degree distribution is not stable, and decays as the rank increases.  Methods of matching a program to specific degree distribution would need to take this decay into account.  Both control flow graphs and data dependency graphs have a degree distribution which follows a power law exponent, and that is higher that of $\gamma=3$, a  measurement which holds across the data collected.  From this we can draw the conclusion that control flow and data dependency networks have Scale-Free properties \cite{li2005towards}.

Figure 3 shows a comparison of data dependency graphs.  This figure shows a sample of data dependency graphs taken from a malicious program with the largest number of segments.  A program was segmented into basic blocks, and a data dependency network was constructed for each basic block.  When analyzing data dependency graphs for a program, the result is a large number of small networks, one per each block segment.  The graphs have been ordered by number of nodes for the networks with the largest size.  It is important to note that degree distributions for the data dependency graphs also follow a power law, and therefore we have not computed the mean degree.  Shown is the network size with respect to number of nodes $N$, the number of edges between nodes or links $L$, the maximum degree $k_{max}$, the ratio of $k_{max}$ to $ln(N)$, the Pearson degree correlation between nodes in the network, and the power law exponent of the degree distribution $\gamma$.  The variable $\gamma$ was calculated by finding a function approximating the cumulative distribution function for the data collected of node degrees.  This distribution of node degrees was used to approximate the cumulative distribution function, and is used for the degree distribution exponent.  In this figure we can see that each network has a cumulative distribution that follows a power law exponent.  From this we can conclude the existence of the Scale-Free property.  We can see that each network has a negative Pearson correlation value, from which we can make an inference about the degree correlation and assortativity.  We can see that the value of $k_{max}$ / $ln(N)$ is low, and this indicates a small network diameter.  This in combination with a high clustering coefficient demonstrates the existence of the Small-World property \cite{alstott2014powerlaw}. 

In Figure 4, $k_{max}$ / $ln(N)$ is a predicted network diameter.  If we assume that a network is Scale-Free, and not a random network, then we would expect the existence of network hubs.  Random networks do not follow a hub-and-spoke pattern, so the existence of hubs would also demonstrate a Scale-Free network.  So the ratio of $k_{max}$ to $ln(N)$ is one metric of a Scale-Free network.  We would expect the network diameters to be very small.  We would also expect the network diameters to decrease as the size of the network grows.  If hubs were not present, then as the size of the network grows, the diameter of the network would grow as well.  The network's diameter is logarithmically dependent upon the network size.  $ln <k>$ is typically used as a metric of network density to demonstrate both Scale-Free and Small-World properties.  Since our networks follow a power law, measured by the exponent $\gamma$, we do not base our measurements on mean degree here.  Instead, if we plot the network diameters in Figure 4, we see that the predicted diameters of the network with respect to $k_{max}$ and $ln(N)$ decrease as the network size $N$ increases.  $k_{max} / ln(N)$ also gives us a prediction of the size of the hub in the network, a result which would not be present in random networks.  We can also see the effect of the hub's presence on the network diameter as $N$ increases \cite{barabasi2013network}.

Since a diameter of path length greater than or equal to 4 does not exist, we can derive the conclusion that any two randomly selected data dependencies will be less than a path distance of 3 nodes away.  Interestingly this appears to hold only for data dependency networks, a property that does not hold for program control flow graphs with large $N$.  From this figure we can also see that small networks have higher density, with the highest degree node taking up a larger proportion of the total network size.  The overall density decreases as the size of the data dependency networks increases.

Figure 5 shows the frequency of network sizes in a frequency histogram of data dependency graphs.  This figure shows the number of nodes in a network, $N$, for data dependency graphs in a program sample.  This shows a power law distribution where most networks are very small, N less than 5.  A small number of networks have a very high number of nodes in the network.  Since this is the distribution, it is not suitable to take the arithmetic mean of the node sizes for the dataset.  The networks for control flow graphs and data dependency graphs both follow power law distributions for their network size.

\subsection{Data Dependency Networks are Degree Disassortative}  Figure 6 shows a scatter plot of data dependency networks.  The network size $N$ is shown on the x-axis, and the Pearson degree correlation is shown on the y-axis.  This graph shows that regardless of network size, the degree correlation coefficient is negative for a large majority of the networks.  While graphs exist with positive Pearson correlation values, no graph with a network size greater than 5 nodes has a Pearson correlation over 0.2.  This indicates the complete absence of graphs of significant size with high Pearson correlation values.  A low or negative correlation coefficient of network degrees indicates that these networks are degree disassortative, and show a preference for not connecting to nodes with high degrees.  This is the case for the sample being analyzed.  This appears to hold only for data dependency networks, and does not hold for control flow graphs with large $N$. 

Through this finding we are able to make the prediction that nodes with high degree will not connect to other similar nodes with high degrees, but show a preference to connect to nodes with low degrees, as shown by the degree correlation value.  This is a structural feature that is less than a purely random network with a degree correlation between hubs based on pure probability.  Since $k_{max}$ can still be a very high degree value, the network topology resulting from this structure is one of a $hub-and-spoke$ pattern, where many nodes with small degree must connect to one of a small set of high degree nodes acting as hubs, and hubs will have fewer connections between each other \cite{barabasi2013network}.

Figure 7 shows a comparison of program control flow graph properties.  These include the degree distribution power law exponent.  The structural properties measured here provide additional structure to the representation of control flow graphs that can be used for further analysis.  Figure 7 shows $k_{max}$ / $ln(N)$ as a measurement of network density and diameter for control flow graphs, and these values are typically very high.  This metric of network diameter is one indication that hubs are not present, and that random networks are accurate models.  We can expect the number of links between nodes in the network to be high, and for individual nodes to have high degrees.  Figure 7 also shows the Pearson correlation values with the network sizes for data dependency graphs.  This shows that a majority of the networks have a negative correlation, and this holds as the network size increases.  The Pearson correlation values for node degrees are very low.  This shows that nodes in a control flow graph do not show a preference for connecting to nodes with high degree, nor do they show a preference for connecting to nodes with low degree.  From this we can draw the conclusion that control flow graphs have Neutral Degree Assortativity.  This assortativity property appears to hold across samples of program control flow graphs.  Since this is the case, random graph models are likely to display the behavior of nodes in this network well through growth and preferential attachment properties.

\begin{figure*}[t]
    \centering
    \hspace{4cm}
    \begin{tabular}{||c|c c c c c c c||} 
     \hline
Malware Sample CFG & $N$ & $L$ & $k_{max}$ & $k1$ & $k2$ & $Pearson$ & $\gamma$ \\ [0.5ex]
     \hline\hline
     \hline
Win32\_APT28\_SekoiaRootkit & 1,495 & 2,779 & 246 & 33.653 & 5.566 & -0.098 & 6.204 \\
Win32\_AgentTesla & 21,732 & 18,394 & 578 & 57.877 & 18.971 & -0.057 & 11.539 \\
Win32\_Avatar & 928 & 1,669 & 23 & 3.366 & 5.337 & -0.012 & 5.999 \\
Win32\_BigBangA & 57,344 & 120,007 & 2,308 & 210.644 & 7.653 & -0.042 & 2.211 \\
Win32\_BigBangB & 46,937 & 97,470 & 2,288 & 212.707 & 7.554 & -0.041 & 2.281 \\
Win32\_BigBangC & 71,109 & 155,022 & 1,153 & 103.204 & 7.587 & -0.054 & 2.250 \\
Win32\_Boaxxe.BB & 2,507 & 5,129 & 118 & 15.076 & 5.555 & -0.073 & 3.618 \\
Win32\_Caphaw\_ShylockA & 1,929 & 3,450 & 76 & 10.046 & 5.935 & -0.046 & 5.934 \\
Win32\_Caphaw\_ShylockB & 1,713 & 3,336 & 45 & 6.043 & 5.476 & 0.038 & 8.291 \\
Win32\_Cridex & 1,155 & 1,386 & 58 & 8.224 & 8.054 & -0.040 & 6.713 \\
Zeus\_Gameover\_2014\_partA & 22,169 & 42,845 & 712 & 71.154 & 7.409 & -0.033 & 2.595 \\
Zeus\_Gameover\_2014\_partB & 20,488 & 39,836 & 599 & 60.336 & 7.310 & -0.039 & 2.544 \\
    %  Zeus\_Banking\_Version\_26Nov2013 & 65 & 67 & 17 & 4.072 & 5.770 & 0.025 & 4.860 &
    %  Win32\_Zeus\_VM & 10 & 7 & 4 & 1.737 & 6.843 & 0 & 5.808 &
    %  Win32\_APT28\_SekoiaRootkit & 102 & 175 & 24 & 5.189 & 3.751 & -0.167 & 2.591 &
    %  Stuxnet\_BDuqu\_Realtek\_partA & 50 & 51 & 7 & 1.789 & 5.487 & 0.210 & 4.11 &
    %  Stuxnet\_BDuqu\_Realtek\_partB & 100 & 162 & 11 & 2.388 & 3.917 & -0.077 & 6.230 &
    %  Stuxnet\_BDuqu\_Realtek\_partC & 107 & 181 & 13 & 2.782 & 3.833 & -0.059 & 7.317 &
    %  Stuxnet\_BDuqu\_Realtek\_partD & 120 & 176 & 9 & 1.879 & 4.166 & 0 & 2.817 &
     \hline
    \end{tabular}
    \caption{Control Flow Graph metrics.  $k1$ and $k2$ represent $k_{max} / ln(N)$ and $ln(N)/ln <K>$ respectively.  $k_{max} / ln(N)$ is large because $k_{max}$ is large.  $ln(N)/ln <K>$ indicates a small world property for large $N$.  $\gamma$ shows the presence of the Scale-Free property.  Control Flow Graphs can have a high number of nodes $N$, and has a high number of links $L$, but still have low Pearson correlation, indicating that their degree assortativity is neutral.}
\end{figure*}

An analysis of adjacent k-cliques would provide a measurement of the hubs and components that are fully connected, and does not include node communities with less than maximum density.  The distribution of adjacent k-cliques in various malware samples show the prevalence of several small communities.  For example, the ZeusGameover\_Feb2014 control flow graph shows the number of 3-cliques to be 3113, the number of 4-cliques to be 31, the number of 6-cliques to be 1 \cite{barabasi2013network}, \cite{palla2005uncovering}.

Other types of networks such as semantic networks have been analyzed and shown to have the Scale-Free and Small-World properties.  It is of note that the networks we have studied have these properties in addition to being degree disassortative \cite{griffiths2007topics}, \cite{steyvers2005large}.

\section{Conclusion}
In this study we have shown the measurement and quantitative analysis of several networks which compose structural features of malicious programs.  In this study we have found through empirical observations that data dependency graphs and control flow graphs in programs are Scale-Free.  DDG and CFG networks have a power law degree distributions, and the degree distribution of control flow graphs is not stable on a log scale and decays with network size.  CFG networks have high diameters, which indicates the absence of hubs.  DDG networks have low diameter, and follow a hub-and-spoke pattern.  CFG network assortativity is neutral and nodes are connected based on a probability distribution.  The distribution of network sizes is positively skewed and follows a power law distribution.  DDG networks correspond to the Small-World property outlined by Watts and Strogatz.  Degree correlations of DDG networks are structurally disassortative.  While DDG nodes are connected to hubs, hubs show a preference for not connecting to similar high degree nodes, and most path lengths between nodes in a data dependency graph are very small.  Since control flow graphs show low correlation and high diameter, random graph models are likely to be better models through modeling growth and preferential attachment of program control flow.  These network properties show that while measurements are skewed, the networks have identifiable structural properties based on degree assortativity and probability.  This serves to provide quantitative analysis and additional structure to the representation of malicious programs for static analysis that can be used for further insights.  The structural properties outlined provide increased feature resolution.  In future studies we intend to use the network features discussed for supervised learning to train models for classification tasks and to more accurately identify patterns of malicious programs correlated to operational semantics.

\subsection{Acknowledgements}
This research was supported in part by Air Force Research Lab grant \#FA8650 to the University of Cincinnati.

Conflicts of Interest: The authors declare no conflicts of interest.

\bibliographystyle{unsrtnat}
\bibliography{references}  %%% Uncomment this line and comment out the ``thebibliography'' section below to use the external .bib file (using bibtex) .





% \end{thebibliography}

\end{document}